\documentclass[twocolumn,amsmath,amssymb]{snp}
\usepackage{graphicx}
\usepackage{dcolumn}
\usepackage{bm}
\begin{document}

\title{{Multiplicity dependent studies for strangeness production with ALICE}}

\author{\large Meenakshi Sharma (for the ALICE Collaboration)$^1$}
\email{meenakshi.sharma@cern.ch}

\affiliation{$^1$Department of Physics, University of Jammu, INDIA}
\maketitle

%

\newcommand{\pp}           {pp\xspace}
\newcommand{\ppbar}        {\mbox{$\mathrm {p\overline{p}}$}\xspace}
\newcommand{\XeXe}         {\mbox{Xe--Xe}\xspace}
\newcommand{\PbPb}         {\mbox{Pb--Pb}\xspace}
\newcommand{\pA}           {\mbox{p--A}\xspace}
\newcommand{\pPb}          {\mbox{p--Pb}\xspace}
\newcommand{\AuAu}         {\mbox{Au--Au}\xspace}
\newcommand{\dAu}          {\mbox{d--Au}\xspace}
\newcommand{\CuCu}         {\mbox{Cu--Cu}\xspace}
\newcommand{\AAa}         {\mbox{A--A}\xspace}

\newcommand{\s}            {\ensuremath{\sqrt{s}}\xspace}
\newcommand{\snn}          {\ensuremath{\sqrt{s_{\mathrm{NN}}}}\xspace}
\newcommand{\pt}           {\ensuremath{p_{\rm T}}\xspace}
\newcommand{\meanpt}       {$\langle p_{\mathrm{T}}\rangle$\xspace}
\newcommand{\alphQCD}{$\alpha_{\mathrm{QCD}}$\xspace}
\newcommand{\alphS}{$\alpha_{\mathrm{s}}$\xspace}
\newcommand{\alphEM}{$\alpha_{\mathrm{em}}$\xspace}
\newcommand{\ycms}         {\ensuremath{y_{\rm CMS}}\xspace}
\newcommand{\ylab}         {\ensuremath{y_{\rm lab}}\xspace}
\newcommand{\etarange}[1]  {\mbox{$\left | \eta \right |~<~#1$}}
\newcommand{\yrange}[1]    {\mbox{$\left | y \right |~<$~0.5}}
\newcommand{\dndy}         {\ensuremath{\mathrm{d}N_\mathrm{ch}/\mathrm{d}y}\xspace}
\newcommand{\dndeta}       {\ensuremath{\mathrm{d}N_\mathrm{ch}/\mathrm{d}\eta}\xspace}
\newcommand{\avdndeta}     {\ensuremath{\langle\dndeta\rangle}\xspace}
\newcommand{\dNdy}           {\ensuremath{\mathrm{d}N_\mathrm{ch}/\mathrm{d}y}\xspace}
\newcommand{\dNdyy}         {\ensuremath{\mathrm{d}N/\mathrm{d}y}\xspace}
\newcommand{\Npart}        {\ensuremath{N_\mathrm{part}}\xspace}
\newcommand{\Ncoll}        {\ensuremath{N_\mathrm{coll}}\xspace}
\newcommand{\dEdx}         {\ensuremath{\textrm{d}E/\textrm{d}x}\xspace}
\newcommand{\RpPb}         {\ensuremath{R_{\rm pPb}}\xspace}
\newcommand{\RAA}         {\ensuremath{R_{\rm AA}}\xspace}

\newcommand{\nineH}        {$\sqrt{s}~=~0.9$~Te\kern-.1emV\xspace}
\newcommand{\seven}        {$\sqrt{s}~=~7$~Te\kern-.1emV\xspace}
\newcommand{\twoH}         {$\sqrt{s}~=~0.2$~Te\kern-.1emV\xspace}
\newcommand{\twosevensix}  {$\sqrt{s}~=~2.76$~Te\kern-.1emV\xspace}
\newcommand{\five}         {$\sqrt{s}~=~5.02$~Te\kern-.1emV\xspace}
\newcommand{\twosevensixnn}{$\sqrt{s_{\mathrm{NN}}}~=~2.76$~Te\kern-.1emV\xspace}
\newcommand{\fivenn}       {$\sqrt{s_{\mathrm{NN}}}~=~5.02$~Te\kern-.1emV\xspace}
\newcommand{\LT}           {L{\'e}vy-Tsallis\xspace}
\newcommand{\GeVc}         {Ge\kern-.1emV/$c$\xspace}
\newcommand{\MeVc}         {Me\kern-.1emV/$c$\xspace}
\newcommand{\TeV}          {Te\kern-.1emV\xspace}
\newcommand{\PeV}          {Pe\kern-.1emV\xspace}
\newcommand{\GeV}          {Ge\kern-.1emV\xspace}
\newcommand{\MeV}          {Me\kern-.1emV\xspace}
\newcommand{\GeVmass}      {Ge\kern-.2emV/$c^2$\xspace}
\newcommand{\MeVmass}      {Me\kern-.2emV/$c^2$\xspace}
\newcommand{\lumi}         {\ensuremath{\mathcal{L}}\xspace}

\newcommand{\ITS}          {\rm{ITS}\xspace}
\newcommand{\TOF}          {\rm{TOF}\xspace}
\newcommand{\ZDC}          {\rm{ZDC}\xspace}
\newcommand{\ZDCs}         {\rm{ZDCs}\xspace}
\newcommand{\ZNA}          {\rm{ZNA}\xspace}
\newcommand{\ZNC}          {\rm{ZNC}\xspace}
\newcommand{\SPD}          {\rm{SPD}\xspace}
\newcommand{\SDD}          {\rm{SDD}\xspace}
\newcommand{\SSD}          {\rm{SSD}\xspace}
\newcommand{\TPC}          {\rm{TPC}\xspace}
\newcommand{\TRD}          {\rm{TRD}\xspace}
\newcommand{\VZERO}        {\rm{V0}\xspace}
\newcommand{\VZEROA}       {\rm{V0A}\xspace}
\newcommand{\VZEROC}       {\rm{V0C}\xspace}
\newcommand{\Vdecay} 	   {\ensuremath{V^{0}}\xspace}

\newcommand{\ee}           {\ensuremath{e^{+}e^{-}}} 
\newcommand{\pip}          {\ensuremath{\pi^{+}}\xspace}
\newcommand{\pim}          {\ensuremath{\pi^{-}}\xspace}
\newcommand{\kap}          {\ensuremath{\rm{K}^{+}}\xspace}
\newcommand{\kam}          {\ensuremath{\rm{K}^{-}}\xspace}
\newcommand{\pbar}         {\ensuremath{\rm\overline{p}}\xspace}
\newcommand{\kzero}        {\ensuremath{{\rm K}^{0}_{\rm{S}}}\xspace}
\newcommand{\lmb}          {\ensuremath{\Lambda}\xspace}
\newcommand{\almb}         {\ensuremath{\overline{\Lambda}}\xspace}
\newcommand{\Om}           {\ensuremath{\Omega^-}\xspace}
\newcommand{\Mo}           {\ensuremath{\overline{\Omega}^+}\xspace}
\newcommand{\X}            {\ensuremath{\Xi^-}\xspace}
\newcommand{\Ix}           {\ensuremath{\overline{\Xi}^+}\xspace}
\newcommand{\Xis}          {\ensuremath{\Xi^{\pm}}\xspace}
\newcommand{\Oms}          {\ensuremath{\Omega^{\pm}}\xspace}
\newcommand{\degree}       {\ensuremath{^{\rm o}}\xspace}
\newcommand{\kstar}        {\ensuremath{\rm {K}^{\rm{* 0}}}\xspace}
\newcommand{\phim}        {\ensuremath{\phi}\xspace}
\newcommand{\pik}          {\ensuremath{\pi\rm{K}}\xspace}
\newcommand{\kk}          {\ensuremath{\rm{K}\rm{K}}\xspace}
\newcommand{\kskm}{$\mathrm{K^{*0}/K^{-}}$}
\newcommand{\phikm}{$\mathrm{\phi/K^{-}}$}
\newcommand{\phixi}{$\mathrm{\phi/\Xi}$}
\newcommand{\phiom}{$\mathrm{\phi/\Omega}$}
\newcommand{\xiphi}{$\mathrm{\Xi/\phi}$}
\newcommand{\omphi}{$\mathrm{\Omega/\phi}$}
\newcommand{\kstf} {K$^{*}(892)^{0}~$}
\newcommand{\phf} {$\mathrm{\phi(1020)}~$}
\newcommand{\dd}{\ensuremath{\mathrm{d}}}
\newcommand{\mT}{\ensuremath{m_{\mathrm{T}}}\xspace}
\newcommand{\krr}{\ensuremath{\kern-0.09em}}
\section*{Introduction}

\par
\begin{figure}
	\includegraphics[width=48mm]{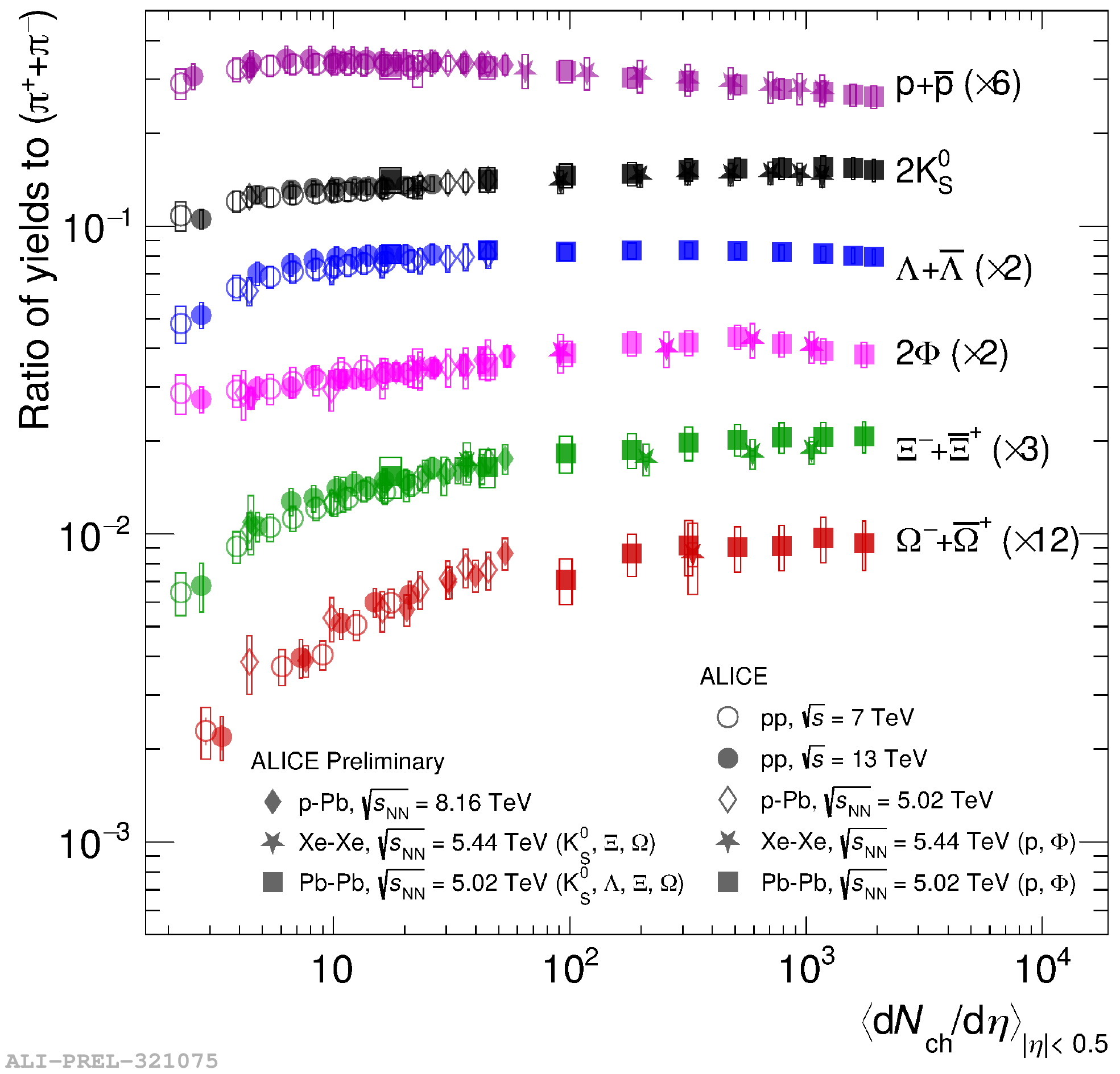}
	\caption{\label{fig1} Ratio of hadron yields to pion yields in different collision systems at different collision energies.}
\end{figure}

 The ALICE experiment has studied strangeness production in different collision systems (pp,~p$-$Pb,~Xe$-$Xe and~Pb$-$Pb) and energies.~Figure \ref{fig1} shows the ratio of the strange particle yield to pion yield as a function of multiplicity for different collision energies and systems. The ratios follow a continuously increasing trend (``enhancement") from low multiplicity pp to high-multiplicity Pb$-$Pb collisions, independent of the initial collision energy and colliding particles. It is also observed that the enhancement is larger for the particles with larger strangeness content.
\par Several other features that were observed in large collision systems and explained as due to collective phenomena related to the formation of the  Quark Gluon Plasma (QGP) are also observed in small systems \cite{a}.
This includes the enhancement of $\Lambda/\mathrm{K^{0}_{S}}$ yield ratios at intermediate $p_{\mathrm T}$, the evolution of the particle $p_{\mathrm T}$ spectra with multiplicity towards higher multiplicity. These unexpected results have motivated to perform further studies using the ALICE detector for understanding the mechanisms responsible for such behaviour in small systems.
\section*{Analysis Details}
In ALICE, the strange hadrons $\mathrm{K^{0}_{S}}$ and $\Lambda$ are reconstructed from their decay particles in the central pseudorapidity region, exploiting the displaced vertex topology induced by the weak decay. The sub-detectors involved in the analyses presented here include the Inner Tracking System (ITS) and the Time Projection Chamber (TPC) which are used for charged particle tracking and identification. The multiplicity intervals are defined based on the signal amplitude measured in the V0A scintillator. A detailed description of the ALICE detector can be found in \cite{b}. 
\section*{Results}
The measurements of $\mathrm{K^{0}_{S}}$ ($\Lambda$) are performed in the rapidity range $-0.5<y<0$ over the $p_{\mathrm T}$ ranges 0.2 (0.6) $<$ $p_{\mathrm T}$ $<$ 10 GeV/$c$. The yield per unit of rapidity, d\textit{N}/d\textit{y}, is obtained by integrating the measured $p_{\mathrm T}$ spectrum and estimating the yield in the unmeasured region using a Levy-Tsallis fit function.
\begin{figure}
	\includegraphics[width=54mm]{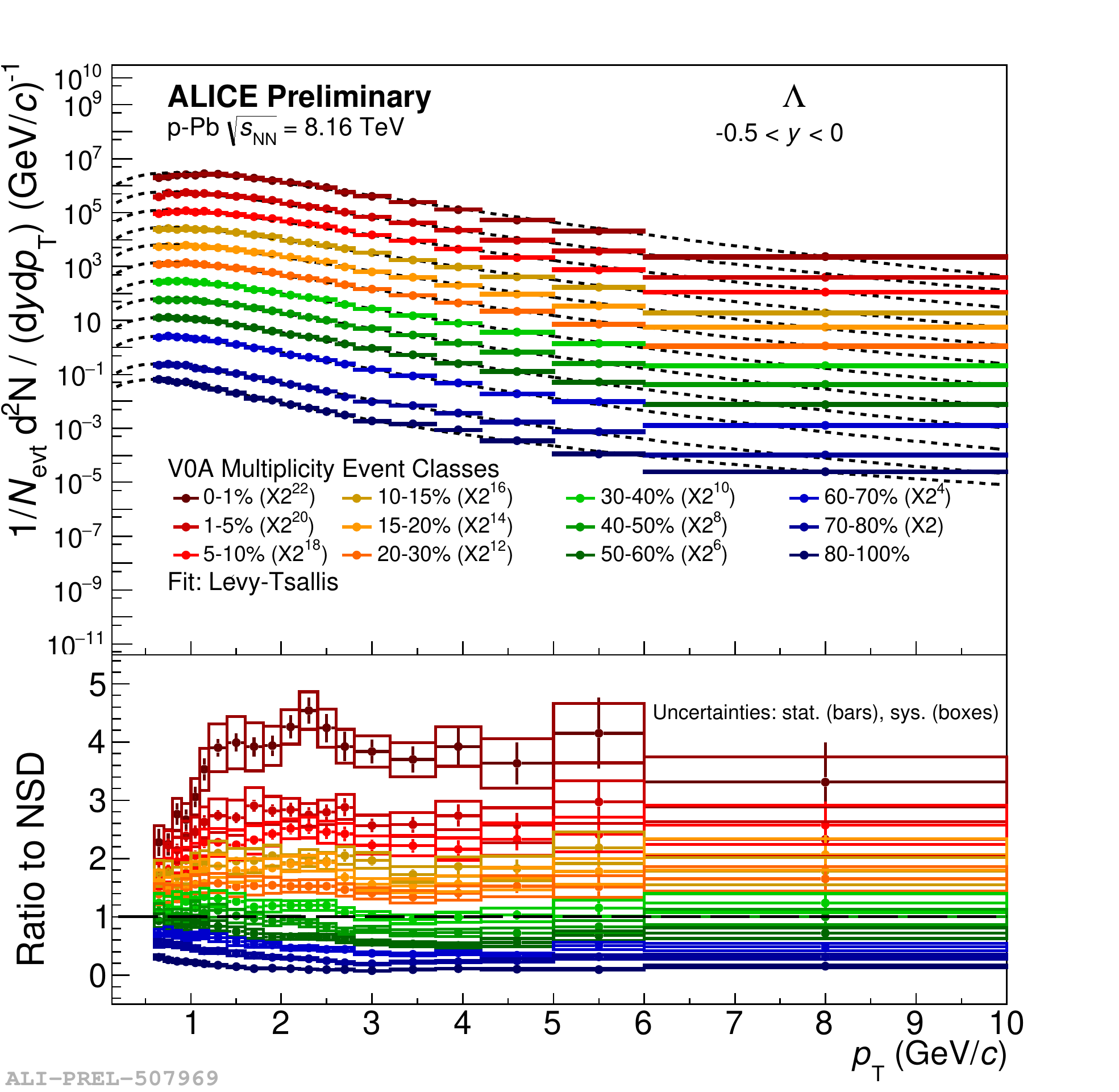}
	\caption{\label{fig2} (Top panel): $p_{\mathrm T}$ spectra of $\Lambda$ in different V0A multiplicity classes. (Bottom panel): Ratio of $p_{\mathrm T}$ spectra in different multiplicity classes to the minimum bias $p_{\mathrm T}$ spectrum.}
\end{figure}
 Figure~\ref{fig2} shows the $p_{\mathrm T}$ spectra for $\Lambda$ in p--Pb collisions at $\sqrt{s_{\mathrm{NN}}}$= 8.16 TeV, in different V0A multiplicity classes along with their ratios to the corresponding spectrum in minimum bias collisions. The ratios suggest that the  $p_{\mathrm T}$ spectra get harder with increasing multiplicity.
 \begin{figure}
 \includegraphics[width=52mm]{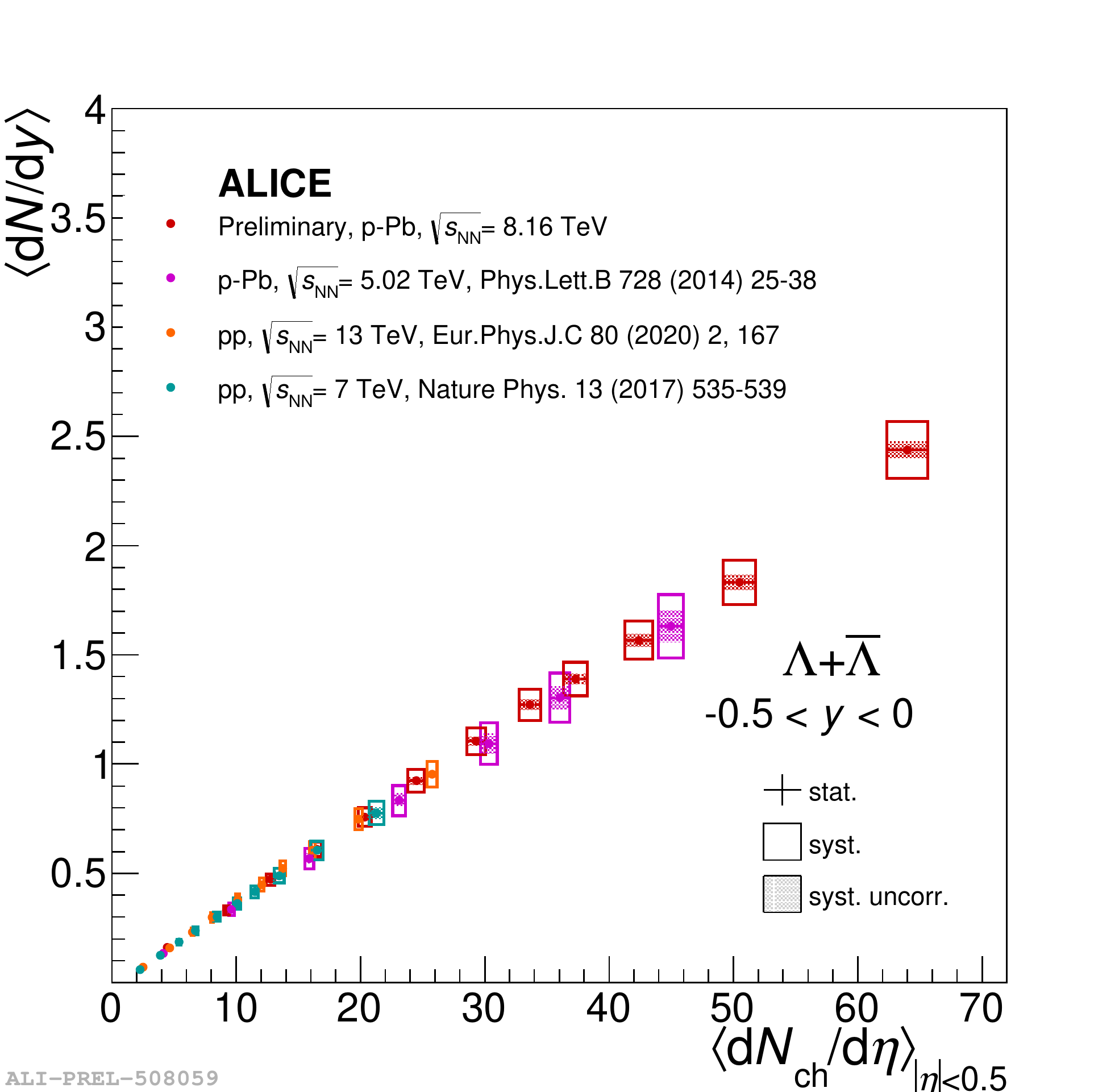}
 \caption{\label{fig3} $\langle d\textit{N}/d\textit{y} \rangle$ of $(\Lambda+\bar{\Lambda})$ as a function of $\langle \mathrm{d}N_\mathrm{ch}/\mathrm{d}\eta \rangle$}
 \end{figure}
 Figure~\ref{fig3} shows the d\textit{N}/d\textit{y} of $\Lambda$ as a function of the average charged particle pseudorapidity density at midrapidity $\langle \mathrm{d}N_\mathrm{ch}/\mathrm{d}\eta \rangle$. These results are compared with the results from different collision systems at different energies. The measurements for different systems are consistent with each other at similar multiplicities and show a common increasing trend. This suggests  that the particle production is independent of collision system and collision energy and mainly driven by $\langle \mathrm{d}N_\mathrm{ch}/\mathrm{d}\eta \rangle$. The $p_{T}$ integrated yield ratio $(\Lambda+\bar{\Lambda})/(\pi^{+}+\pi^{-}$) as a function of $\langle \mathrm{d}N_\mathrm{ch}/\mathrm{d}\eta \rangle$ in different systems is shown in Figure \ref{fig4} (top panel). A smooth evolution is observed for this ratio going from from low to high multiplicity. Moreover, the baryon over meson ratio $\Lambda/\mathrm{K^{0}_{S}}$ in two different multiplicity intervals and energies is shown in Figure \ref{fig4} (bottom panel). A significant peak is observed at intermediate $p_{\mathrm T}$ which can be explained by effects of radial flow combined with processes like recombination during the hadronisation of the created QGP. The measured trend with multiplicity shows a more pronounced peak in high-multiplicity (0--5$\%$) as compared to low multiplicity (60--80$\%$) p--Pb collisions and no dependence on the collision energy within uncertainties.
   \begin{figure}
 \includegraphics[width=52mm]{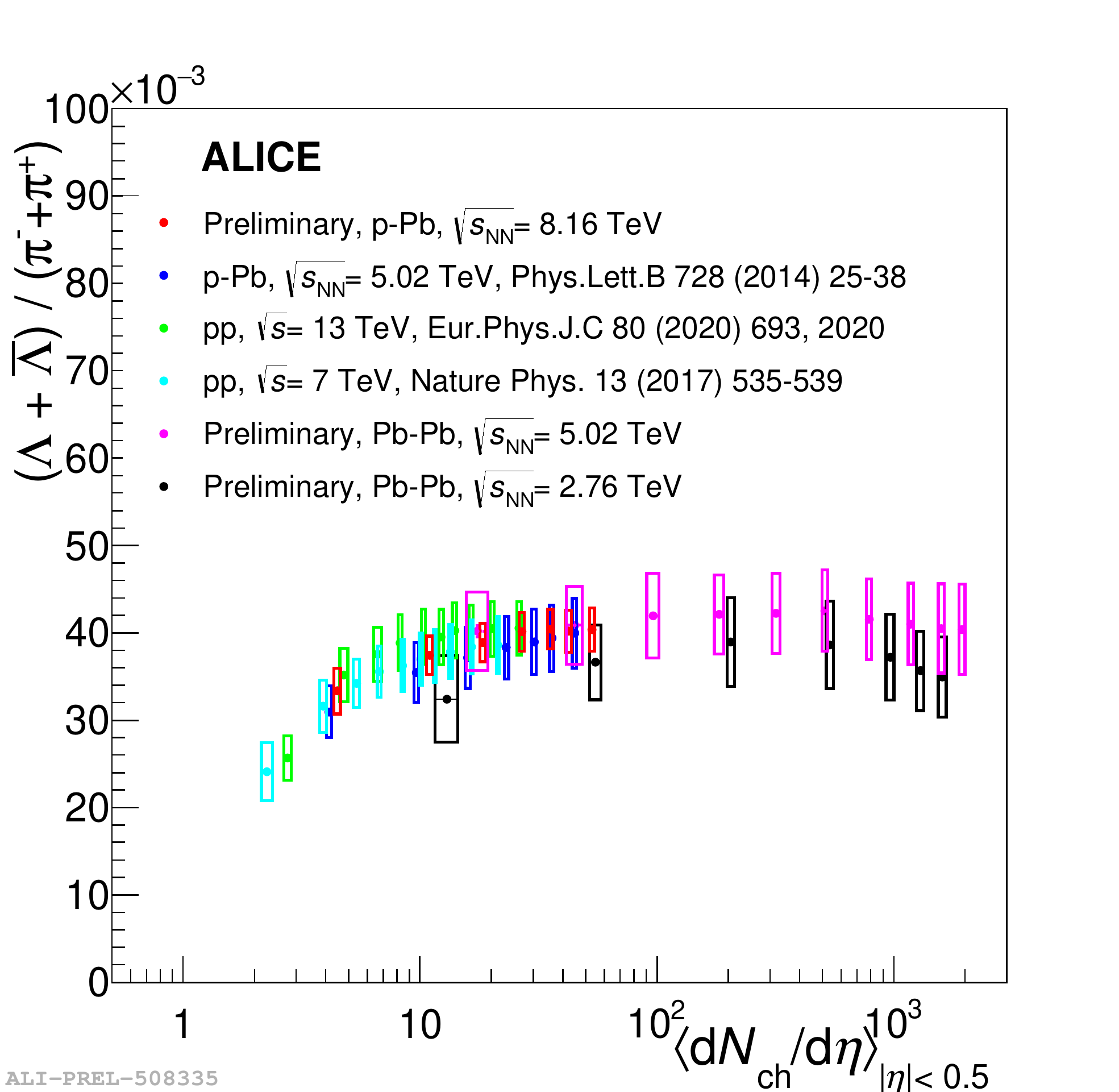}
 \includegraphics[width=52mm]{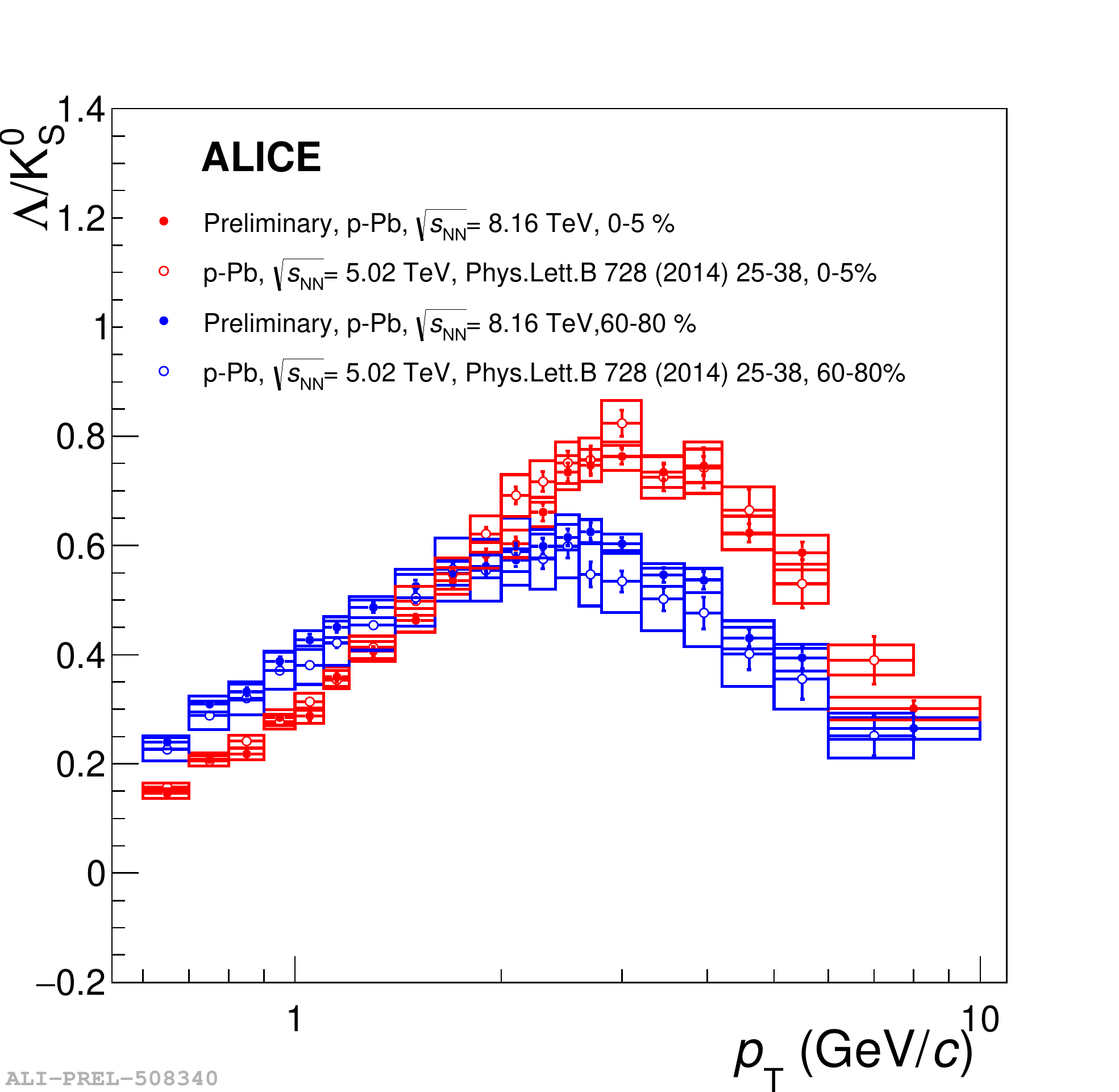}
 \caption{\label{fig4} (Top): $(\Lambda+\bar{\Lambda}) /(\pi^{+}+\pi^{-}$) as a function of $\langle \mathrm{d}N_\mathrm{ch}/\mathrm{d}\eta \rangle$. (Bottom): $\Lambda/\mathrm{K^{0}_{S}}$ as a function of $\langle \mathrm{d}N_\mathrm{ch}/\mathrm{d}\eta \rangle$.}
 \end{figure}

\end{document}